\documentclass[aps,prl,twocolumn,groupedaddress]{revtex4}

\usepackage{graphicx}

\begin{document}

\title{Evolution of the Electronic Structure of 1T-Cu$_{x}$TiSe$_{2}$}

\author{J. F. Zhao$^1$, H. W. Ou$^1$, G. Wu$^2$,  B. P. Xie$^1$, Y. Zhang$^1$, D. W. Shen$^1$,  J.
Wei$^1$,  L. X. Yang$^1$, J. K. Dong$^1$,  M. Arita$^3$, H.
Namatame$^3$, M. Taniguchi$^3$, X. H. Chen$^2$, and
 D.L. Feng$^{1}$} \email{dlfeng@fudan.edu.cn}

\affiliation{$^1$Department of Physics, Applied Surface Physics
State Key Laboratory, and Advanced Materials Laboratory, Fudan
University, Shanghai 200433, P. R. China}

\affiliation{$^2$Hefei National Laboratory for Physical Sciences at
Microscale and Department of Physics, University of Science and
Technology of China, Hefei, Anhui 230026, P. R. China}

\affiliation{$^3$Hiroshima Synchrotron Radiation Center and Graduate
School of Science, Hiroshima University, Hiroshima 739-8526, Japan}

\date{\today}

\begin{abstract}
The electronic structure of a new charge-density-wave/
superconductor system, 1T-Cu$_{x}$TiSe$_{2}$, has been studied by
photoemission spectroscopy. A correlated semiconductor band
structure is revealed for the undoped case. With Cu doping, the
charge density wave is suppressed by the raising of the chemical
potential, while the superconductivity is enhanced by the
enhancement of the density of states. Moreover, the strong
scattering at high doping might be responsible for the suppression
of superconductivity in that regime.
\end{abstract}

\maketitle


Transition metal dichalcogenides (TMD's) provide an important
playground of interesting physics. Different chemical and
structural configurations cause dramatic changes of their
properties. For example, charge density wave (CDW) in two
dimensional (2D) systems was first discovered in
TMD's\cite{WilsonAP}. In the CDW state, an energy gap opens at the
Fermi surface of 1T-structured TaS$_{2}$\cite{ThPillo}, but only
partially opens for the 2H-structured TaS$_{2}$\cite{ShenNaTaS},
whereas there is no gap for 1T-TiSe$_{2}$\cite{Aebi2}. Moreover,
superconductivity usually coexists and competes with CDW in
2H-structured TMD's\cite{Valla2004,Castro,shinScience}, whereas it
rarely exists in 1T structured compounds.

Recently, the discovery of superconductivity in
1T-Cu$_{x}$TiSe$_{2}$ has further enriched the physics of TMD's
\cite{cavaNP}. The undoped 1T-TiSe$_{2}$ is a CDW material, whose
mechanism remains controversial after decades of research. For
example, some considered the CDW a band-type Jahn-Teller effect,
where the electronic energy is lowered through structural
distortion\cite{JT1,JT2}. Some considered it  a  realization of
the excitonic CDW mechanism proposed by Kohn in the
1960's\cite{Kohn67,Wilson}; but different models were proposed to
interpret the electronic structure, depending on whether system
was argued to be a semi-metal, or a
semiconductor\cite{Kidd02,Aebi01}. With Cu doping, it was found
that the CDW transition temperature quickly drops, similar to
other M$_x$TiSe$_2$'s (M=Fe,Mn,Ta,V and
Nb)\cite{Cui2006,Salvo78,Levy1980,Baranov2007}. Meanwhile, the
superconducting phase emerges from $x\sim0.04$ and reaches the
maximal transition temperature of 4.3K at $x\sim0.08$, then
decreases to 2.8K at $x\sim 0.10$. Quite remarkably, this phase
diagram resembles those of the cuprate and heavy fermion
superconductors\cite{Sciencereview}, except here the competing
order of superconductivity is the charge order, instead of the
antiferromagnetic spin order. The presence of this ubiquitous
phase diagram in 1T-Cu$_{x}$TiSe$_{2}$  calls for a detailed study
of its electronic structure. In particular, the information
retrieved might help resolve the controversy on the CDW mechanism
for 1T-TiSe$_{2}$.

\begin{figure*}[t!]
\includegraphics[width=17.5cm]{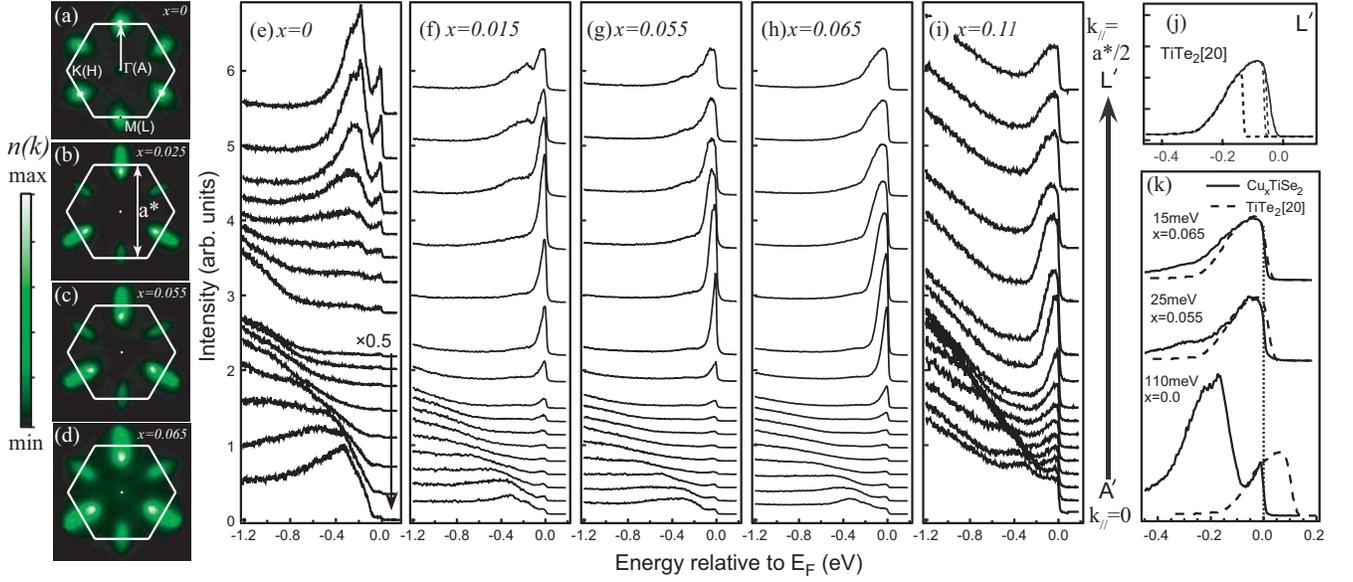}
\caption{(a-d) False color plot of the distribution of spectral
weight integrated over $\pm$40\,meV around $E_F$  for
1T-Cu$_{x}$TiSe$_{2}$ with  $x$=0,  0.025, 0.055, and 0.065
respectively. Data were 3-fold symmetrized. (e-i) Photoemission
spectra sampled equally along the $k_{\parallel}=0 \rightarrow
a^{*}/2$ cut shown in panel \textbf{a} for various Cu dopings. The
spectra were stacked for clarity. (j) Photoemission spectrum of
TiTe$_2$ at \textbf{L${'}$} taken at 20K with 21.2eV photon from
Ref\cite{TiTe2}, where the three dashed curves are the simulated
spectra when the chemical potential is shifted down by 15meV,
25meV, and 110meV respectively (lifetime variation with doping is
neglected). (k) Comparison of the shifted TiTe$_2$ spectra with
the \textbf{L${'}$} spectra for three dopings. All data were taken
at T=20K. } \label{f1}
\end{figure*}


We studied 1T-Cu$_{x}$TiSe$_{2}$ with high resolution angle
resolved photoemission spectroscopy (ARPES). A correlated
semiconductor band structure of the undoped system is evidently
illustrated, resolving a long-standing controversy.  Cu doping is
found to effectively enhance the density of states  around the
Fermi energy ($E_F$), which explains the enhancement of
superconductivity. On the other hand, severe inelastic scattering
was observed near the solubility limit, corresponding to the drop
of superconducting transition temperature in that regime. With
increased doping, chemical potential is raised, and signs of the
weakening electron-hole coupling is discovered, which is
responsible for the suppression of the CDW.  Our results indicate
that the seeming "competition" between CDW and superconductivity
in the phase diagram is a coincidence caused by different effects
of doping in this 1T compound, in contrast to the 2H-TMD
case\cite{ShenNaTaS}.


1T-Cu$_{x}$TiSe$_{2}$  single crystals were prepared by the
vapor-transport technique, with doping $x=0, 0.015, 0.025, 0.055,
0.065$ and $0.11$ (accurate within $\pm 0.005$)\cite{WuGang}. The
copper concentrations were determined by inductively coupled
plasma spectrometer chemical analysis and confirmed by $c$-lattice
parameter calibration\cite{cavaNP}. The superconducting phase
transition temperatures, $T_{c}$'s, for $x=0.055$ and 0.065 are
2.5 and 3.4 K respectively, similar to what was reported
before\cite{cavaNP}. Superconductivity is not observed down to 2K
for $x=0.11$. Judging from the reported phase diagram, the CDW
phase transition temperature $T_{CDW}$'s are about 220K, 190K,
170K and 70K for $x=0, 0.015, 0.025$ and $0.055$ respectively, and
doping $0.065$ is just merely outside the CDW regime. ARPES
experiments were performed with Scienta R4000 electron analyzers
and photons from a Helium gas discharge lamp and synchrotron
radiation at beam line 9 of HiSOR. The angular resolution is
$0.3^\circ$ and the energy resolution is 10\,meV. The samples were
cleaved/measured in ultra-high vacuum ($\sim 5\times
10^{-11}\,mbar$).


Survey of the electronic structure was conducted with 21.2eV
photons from a Helium lamp. The photoemission intensity
distributions near $E_F$ are projected onto the
\textbf{$\Gamma$-K-M} or \textbf{A-H-L} plane for $x=0, 0.025,
0.055$ and $0.065$ in Fig.\,\ref{f1}(\textbf{a}-\textbf{d})
respectively.  The perpendicular momentum of
$k_{\parallel}=a^{*}/2$ electrons is close to that of \textbf{L}
(thus denoted by \textbf{L${'}$}), where $a^{*}$ being the
reciprocal lattice vector shown in
Fig.1\textbf{b}\cite{Aebi01,kzeffects}. Strong spectral weight is
located around  \textbf{L${'}$}, forming the so called Fermi
patches as observed in cuprate superconductors and some 2H-TMD
compounds\cite{ShenNaTaS}. The Fermi patches expand with increased
Cu doping, consistent with the rising carrier density, and the
susceptibility data\cite{cavaNP}. Symmetry of the triagonal
lattice is manifested in the alternating strong and weak
\textbf{L${'}$} regions. As they show similar behaviors, we will
focus on the data around the strong region hereafter.


Photoemission spectra taken at low temperatures along the cut
sketched in Fig.1\textbf{a} are shown in
Fig.\,\ref{f1}\textbf{e}-\textbf{i} for systems ranging from the
undoped to the Cu solubility limit. The spectra exhibit remarkable
doping dependence. The flat-band feature at $E_F$ around
\textbf{L${'}$} corresponds to the narrow Ti $3d$ band, which
dominates the density of states near $E_F$. The highly dispersive
features near $k_{\parallel}=0$ are spin-orbit-split Se $4p$
bands\cite{Kidd02,Aebi01},  which are folded to the \textbf{L}
region (Fig.\,1\textbf{e-g}) as will be discussed below. The Ti
$3d$ band clearly extends from around the \textbf{L${'}$} region
towards $k_{\parallel}=0$ with increased doping, causing
significant increase of the density of states around $E_F$. Recent
thermal conductivity data from 1T-Cu$_{0.06}$TiSe$_{2}$ indicate
that it is a single band $s$-wave superconductor\cite{Lee}.
Therefore, if the superconductivity here is of BCS kind as in the
2H-TMD compounds\cite{shinScience,Valla2004},  increased Cu doping
would certainly enhance it. However, when the Ti $3d$ band
eventually reaches  $k_{\parallel}=0$ at $x=0.11$
(Fig.\,1\textbf{i}), the spectral background at high binding
energies is also severely enhanced, which has been confirmed in
different batches of samples. We note that the residual
resistivity ratio of $x=0.11$ sample is similar to the
others\cite{WuGang}. Therefore, it suggests that high Cu
concentrations would induce disorder effects such as enhanced
inelastic scattering, which might be responsible for the
suppression of superconductivity in this regime \cite{cavaNP}. On
the other hand, backward scattering that would enhance the
residual resistivity seems not to be affected to the same extent.
To fully understand these effects, local measurements, such as
scanning tunnelling microscopy, may be necessary to provide a more
direct picture.


The spectral linewidth of the Ti $3d$ band shows an intriguing
increase with doping at the same momentum, while the mid-point of
the spectral leading edge is about 3$\sim$4\,meV above $E_F$. In a
conventional band structure sense, this would indicate the
``quasiparticle" band has not crossed $E_F$ up to the highest Cu
doping. The observed sharp feature at low doping is just a
Fermi-Dirac cut-off of a broad feature above $E_F$. Therefore, it
grows broader with increasing doping, and eventually the peaks
become rounded or even flat(Fig.\,\ref{f1}\textbf{h}-\textbf{i}).
This remarkably resembles the lineshape of TiTe$_2$ spectrum taken
under the same condition\cite{TiTe2}, as reproduced in
Fig.\,\ref{f1}\textbf{j}. TiTe$_2$ is a metal, whose Ti-$3d$-band
feature at \textbf{L${'}$} is just below $E_F$. In fact, if one
would shift the chemical potential down by 15\,meV, 25\,meV, and
110\,meV (with an uncertainty of about $\pm 10$\,meV)
(Fig.\,\ref{f1}\textbf{j}), or equivalently shift the  TiTe$_2$
spectra up (Fig.\,\ref{f1}\textbf{k}), the 1T-Cu$_{x}$TiSe$_{2}$
($x=0.065, 0.055$ and $0$ respectively)  spectra would be fitted
surprisingly well. The differences at higher energies between the
shifted spectra and the data are attributed to the folded Se $4p$
bands due to the CDW, as will be discussed below. TiTe$_2$ was
considered to be a prototypical Fermi liquid system before. With
much improved resolution nowadays, the many-body nature of
TiTe$_2$ and 1T-Cu$_{x}$TiSe$_{2}$ is exposed by the broad and
non-Lorentzian lineshape, and its unconventional doping
dependence. In particular, our fitting shows that the sharp Ti
$3d$ feature at \textbf{L${'}$} for $x=0$ is just a partially
occupied spectral function, whose center-of-mass is actually
\textit{above} the Fermi energy, whereas most of the spectrum is
recovered for $x=0.065$, since it differs very little from the
1T-Cu$_{0.11}$TiSe$_{2}$  or TiTe$_2$ spectrum.

\begin{figure}[t]
\includegraphics[width=8.5cm]{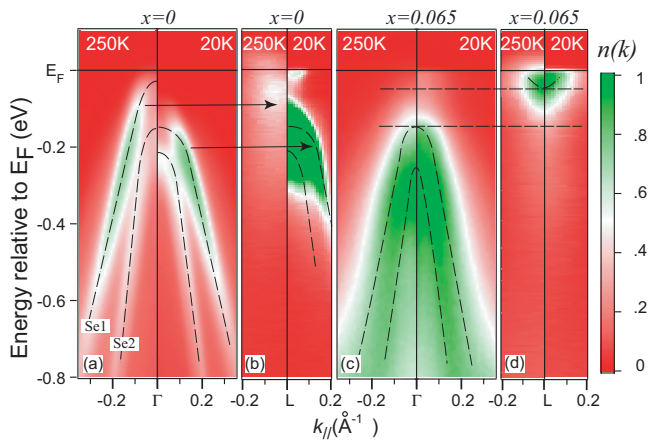}
\caption{Photoemission intensity map for (a-b) 1T-TiSe$_{2}$ and
(c-d) 1T-Cu$_{0.065}$TiSe$_{2}$  at 20K and 250K were compared
near $\Gamma$ and \textbf{L}. The dash-dotted line is the
dispersion obtained from the MDC analysis. Arrows illustrate the
band folding. Se1 and Se2 labels are assigned to the two Se $4p$
bands as illustrated.} \label{Comp}
\end{figure}

The anomalous evolution of the Ti $3d$ state lineshape requires
reexamination of the band structure. Photoemission intensity along
the $\Gamma$-\textbf{M}, and \textbf{L}-\textbf{A} directions were
measured with 12.85\,eV and 18.5\,eV photons
respectively\cite{Kidd02} (Fig.2). Data at both low and high
temperatures were compared.  For 1T-TiSe$_{2}$, at 250K (above the
CDW transition), the top of the valence band is located slightly
below the Fermi energy (Fig.2\textbf{a}), while the Ti $3d$ band
feature is quite weak and broad around $E_F$(Fig.2\textbf{b}).  At
20K, a large gap opens near $\Gamma$ region, and the Se $4p$ bands
are strongly folded to \textbf{L} region due to the formation of
$2\times2\times2$ CDW.\cite{Kidd02}. The CDW fluctuations are
strong even in the normal state, which induces band folding and
possible small gap at $\Gamma$. Since thermal broadening above the
CDW transition temperature further adds to the complication,
whether the Se1 band and the Ti band overlap or not could not be
definitely resolved, resulting in a controversial
situation\cite{Aebi01,Kidd02}. Fortunately,
1T-Cu$_{0.065}$TiSe$_{2}$ provides an excellent opportunity to
resolve this controversy. As shown in Fig.2\textbf{c-d}, the band
position hardly moves at all with temperature in this compound as
it is already out of the CDW phase. CDW fluctuations still exist,
but only causes quite weak band folding. Therefore, the original
band structure without much complication from the CDW and thermal
broadening effects is revealed. Furthermore, as chemical potential
shifts up with electron doping, the Ti $3d$ feature at \textbf{L}
is almost fully recovered (Fig.2\textbf{d}). There is a clear gap
as large as 110\,meV between the Se1 and Ti $3d$ bands, whose
position is defined by the centroid of the spectra. The undoped
system is thus \textit{undoubtedly} a semiconductor under this
conventional definition. On the other hand, we emphasize that it
is a correlated semiconductor, and the broad ``tails" of those two
bands could overlap in-between the two dashed lines in
Fig.2\textbf{c-d}. In the undoped case, these tails induces holes
near $\Gamma$, and electrons near \textbf{L}, giving the system a
weak semi-metallic nature in the many-body sense.


\begin{figure}[t]
\includegraphics[width=8cm]{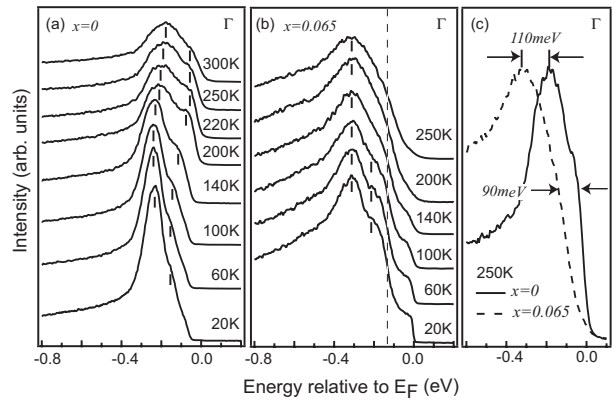} \caption{ Temperature
dependence of the spectrum at $\Gamma$ for (a) 1T-TiSe$_{2}$, and
(b) 1T-Cu$_{0.065}$TiSe$_{2}$.  (c) compares the spectral leading
edges of the Se $4p$ band at $\Gamma$ for these two dopings at
250K. } \label{f4}
\end{figure}

Finer temperature evolution of the spectrum at $\Gamma$  is shown
in Fig.3. For 1T-TiSe$_{2}$,  a large gap opens at the Se
4\emph{p} bands when the system enters deeply into the CDW states
(Figs.3\textbf{a}), which causes a shift of about  110 meV and 50
meV for the Se1 and Se2 band respectively. This helps lowering the
energy of the CDW state. For $x =0.065$, such a shift is
absent(Fig.3\textbf{b}), and there is only temperature broadening
of the step-like feature at E$_{F}$, which comes from the
extension of the Ti 3d bands as illustrated in Fig.1\textbf{h},
plus possibly some weak folding (Fig.2\textbf{c}). Because the
1T-TiSe$_{2}$ spectrum varies very little at high temperatures,
one can estimate a chemical potential shift of about
100$\pm10$\,meV with 6.5\% Cu doping both from the leading edge
and the peak position (Fig.3\textbf{c}). This is consistent with
the simulation in Fig.1\textbf{k}.


The folded Se $4p$ bands at \textbf{L} weaken significantly with
increased doping or temperature, in correlation with the CDW
strength (Fig.4).  CDW fluctuations are quite robust in the normal
state. For $x=0$, the spectrum at 250K still contains significant
contribution from the folding (Fig.\,4\textbf{a}). Similarly in
Fig.\,4\textbf{b-d}, there also exists a small amount of residual
folded weight well above $T_{CDW}$. Moreover, with increased
temperature, the folded features shift toward $E_F$. For example,
a shift of more than 110 meV for the folded Se1 ban is observed in
Fig.\,4\textbf{a} for TiSe$_2$. The positions of the folded Se1
and Se2 bands are summarized in Fig.4\textbf{e}. One finds that
the shift of the Se1 band is much larger than that of Se2 band in
the studied temperature range. With increased doping, the maximal
shift of the Se1 band drops faster than that of the Se2, and
eventually they both diminish with the CDW. The correlation
between the shift and CDW cannot be understood within the band
Jahn-Teller picture, since the Jahn-Teller distortion would have
gained more electronic energy through the filled Ti 3d band. On
the other hand, if the shift of the Se bands are caused by the
coupling between the electrons in the Se bands and holes in the Ti
band (\textit{i.e.} the excitonic effects), the decreasing shift
of the Se bands with doping can be naturally explained by the
weakening electron-hole interaction due to the raise of chemical
potential. As  the energy of the CDW state lowers through these
shifts, the CDW strength would decrease with the weakening
electron-hole interaction. Consistently, the indirect Jahn-Teller
scenario proposed for the undoped system before\cite{Kidd02},
which includes electron-hole interactions, conduction band
degeneracy, and the spin-orbit splitting of the valence bands, is
also able to explain the above doping behavior. In this scenario,
the degenerate conduction bands split and push down the valence
band in the CDW state and thus gain energy\cite{Kidd02}. Since Se1
band is closer to the conduction band, the hybridization will push
it harder than the Se2 band(Fig.4\textbf{e}).

\begin{figure}[t]
\includegraphics[width=8.5cm]{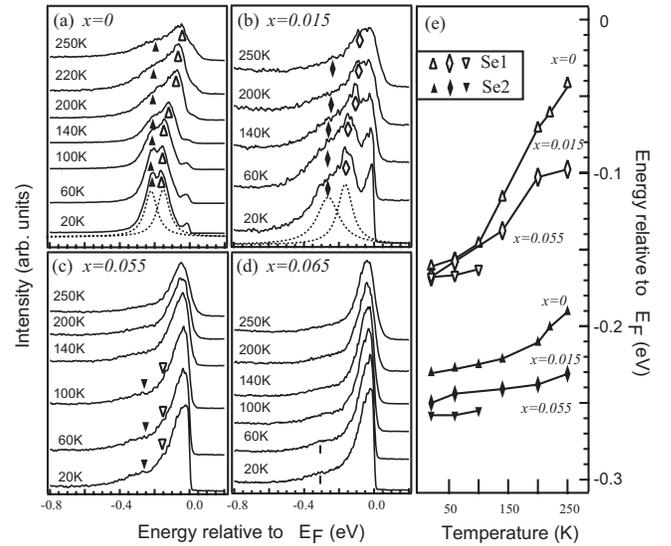}
\caption{ Temperature dependence of the spectrum at L
 for (a) $x=0$, (b) $x=0.015$, (c)
$x=0.055$ and (d) $x=0.065$. The spectra were stacked for clarity.
The dashed lines are fit to help resolve the folded two Se $4p$
bands.  (e) summarizes the temperature dependence of  Se1 and Se2
positions in panels a-d. Error bars of $\pm 5$\,meV is neglected for
a clearer view.}
\end{figure}


Alternatively, the drop of CDW can be understood in the general
excitonic picture proposed by Kohn\cite{Kohn67}, where an
indirect-gap semiconductor is unstable against the formation of
excitons between valence electrons and holes in the conduction
band, if the exciton binding energy is higher than the energy
difference between the electron and hole (\textit{i.e.} band gap
for a semiconductor). For 1T-Cu$_{x}$TiSe$_{2}$, the measured
chemical potential shift of 100$\pm10$\,meV for $x=0.065$ makes
the centroid of the valence band sufficiently below $E_F$ and the
exciton/CDW formation costly enough. Therefore, the disappearance
of CDW at high doping is expected. Compared with the exciton
energy of 17\,meV estimated earlier\cite{Aebi2}, this large shift
is necessary because there is strong interactions in the system
(most likely strong electron phonon
interactions\cite{chiangphonon}), and the large linewidth would
reduce the distance between the highest valence state and $E_F$
significantly.


Compared with other 1T-TMD compounds, the CDW in
1T-Cu$_{x}$TiSe$_{2}$ does not cause an energy gap for the Ti $3d$
band, which makes the large density of states at $E_F$. Therefore,
there is plenty of spectral weight available for
superconductivity. This is perhaps why superconductivity in
1T-TMD's is first discovered here, when large spectral weight near
$E_F$ is induced by Cu doping.  Moreover, our results suggest that
the seeming competition between CDW and superconductivity in this
system is very likely a coincidence, as the doping will increase
the density of states and raise the chemical potential
simultaneously.


To summarize, 1T-Cu$_x$TiSe$_2$ is proved to be a correlated
semiconductor at zero doping, and the doping behavior of the CDW
in this material can be well understood if electron-hole coupling
is taken into account. Cu doping enhances the density of states
and thus favors the superconductivity, while it also raises the
chemical potential and weakens the charge density wave.

We gratefully acknowledge the helpful discussion with Prof. Z. Y.
Weng and Prof. D. H. Lee. This work was supported by NSFC, MOST
(973 project No.2006CB601002 and No.2006CB921300), and STCSM of
China.


\begin{references}
\bibitem{kzeffects} With Cu intercalation, $c$ only
increases by 5\% at the maximum concentration. Our synchrotron
experiments show that the photon energy corresponding to
particular momentum varies negligibly with dopings.
\end{references}
\end{document}